\documentclass[12pt, draftclsnofoot, onecolumn]{IEEEtran}

\usepackage{xcolor,mathtools,subfigure}
\usepackage{algorithm}
\usepackage{algpseudocode}
\usepackage{epsfig,makeidx,color}
\usepackage{amsmath,amssymb,bbm}
\usepackage{cite,graphicx,lipsum}
\usepackage{enumerate}
\usepackage[switch,pagewise]{lineno}
\usepackage{hyperref}
\hypersetup{
        colorlinks = true,
        citecolor=blue,
}
\newcommand{\indicator}{\mathbbm{1}}

\def\uE{{\mathbb E}}

\newtheorem{mylemma}{\bf Lemma} 

\def\be{ \begin{equation} }
\def\ee{ \end{equation} }
\def\bea{ \begin{eqnarray} }
\def\eea{ \end{eqnarray} }

\def\b0{{\bf 0}}

\ifCLASSOPTIONonecolumn
\else

\fi

\ifCLASSOPTIONonecolumn
\fi

\begin{document}

\title{Enhancing Reliability in LEO Satellite Networks via High-Speed Inter-Satellite Links}
\author{Jinho Choi\\
\thanks{The author is with
the School of Information Technology,
Deakin University, Geelong, VIC 3220, Australia
(e-mail: jinho.choi@deakin.edu.au).}}


\maketitle
\begin{abstract}
Low Earth orbit (LEO) satellites play a crucial role in providing global connectivity for non-terrestrial networks (NTNs) and supporting various Internet-of-Remote-Things (IoRT) applications. Each LEO satellite functions as a relay node in the sky, employing store-and-forward transmission strategies that necessitate the use of buffers. However, due to the finite size of these buffers, occurrences of buffer overflow leading to packet loss are inevitable. In this paper, we demonstrate how inter-satellite links (ISLs) can mitigate the probability of buffer overflow. Specifically, we propose an approach to reallocate packets among LEO satellites via ISLs to minimize the occurrence of buffer overflow events. Consequently, the implementation of ISLs can lead to a more reliable satellite network, enabling efficient packet reallocation to reduce the probability of buffer overflow.

\end{abstract}


\section{Introduction}

While terrestrial networks play a crucial role in providing connectivity for devices and things across a wide range of Internet-of-Things (IoT) applications, their coverage might be limited in some areas, including remote or underserved areas. For example, there might be devices and sensors operating in remote regions for Internet-of-Remote-Things (IoRT) applications \cite{Sanctis16}. Consequently, non-terrestrial networks (NTNs) have been considered to ensure global coverage  by the 3rd generation partnership project (3GPP) Release 17 and onward \cite{Giordani21} \cite{Azari22}. Among various technologies for NTNs, low Earth orbit (LEO) satellites have attracted considerable attention \cite{Pachler21}. 

There are numerous LEO satellites orbiting at altitudes ranging from 180 to 2,000 kilometers above the Earth's surface. In each orbit, there are a few tens LEO satellites \cite{Pachler21}.
Due to their lower altitude, LEO satellites can be arranged in denser constellations compared to satellites in higher orbits, enabling LEO mega constellations to provide global coverage with high data rate services \cite{Homssi22}.

While satellites can function as relay nodes with a two-hop transmission, comprising ground user-satellite and satellite-ground gateway links, it is also feasible to establish direct connections between satellites using inter-satellite links (ISLs) \cite{Chaudhry21}. 
While the channel conditions of ISLs are stable, those of ground user-satellite and satellite-ground gateway links can vary due to the orbital movement of LEO satellites and weather conditions \cite{Loo98}.
If neighboring LEO satellites in an orbit are interconnected using ISLs, the resulting network topology forms a ring, enabling efficient packet forwarding from one LEO satellite to another through ISLs. This interconnected arrangement enhances network resilience and reduces reliance on ground-based infrastructure, making it particularly advantageous for remote or inaccessible regions. Furthermore, each LEO satellite can communicate directly with ground users and gateways, resulting in a network configuration with multiple inputs and multiple outputs.

It is important to note that ISLs do not enhance the capacity of an LEO satellite communication network as a multi-input multi-output system. To illustrate this,  consider one-directional traffic flows from ground users to ground gateways via LEO satellites. In such scenarios, regardless of the presence of ISLs, the achievable sum rate depends solely on the capacities of the ground user-satellite and satellite-ground gateway links.

In this paper, we study the probability of buffer overflow in LEO satellite networks equipped with on-board processing (OBP) capability \cite{Ippolito2017} and utilizing store-and-forward transmission strategies. It is crucial to emphasize that LEO satellites rely on store-and-forward transmission to initially store received packets from ground users in buffers before forwarding them to ground gateways or other satellites via ISLs. Moreover, in IoRT applications \cite{Sanctis16}, ground users, such as devices and sensors, may exhibit sporadic activity and send short packets when active. Therefore, in our analysis of packet transmission, we examine the probability of buffer overflow at LEO satellites employing store-and-forward transmission. Given the finite buffer size in such scenarios, it becomes imperative to maintain a low probability of buffer overflow to ensure reliable network performance and prevent packet loss. 
To address this, we propose an approach to reduce the probability of buffer overflow by reallocating packets among LEO satellites via ISLs. The novelty of our proposed approach lies in its effective strategy to optimize buffer management through packet reallocation via ISLs, thereby enhancing network reliability.

\section{System Model}

\begin{figure}[thb]
\begin{center}
\includegraphics[width=0.6\columnwidth]{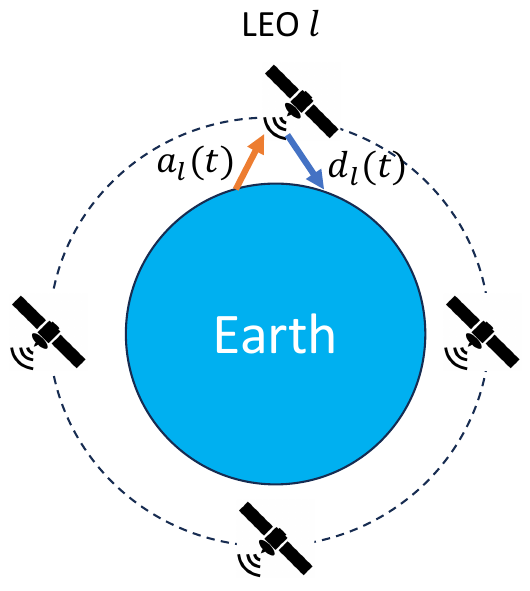} 
\end{center}
\caption{LEO satellites in an orbit.}
        \label{Fig:LEO_ad}
\end{figure}

Suppose there are $L$ LEO satellites in an orbit as shown in Fig.~\ref{Fig:LEO_ad}. 
We assume that each LEO satellite can receive data packets from ground users within its footprint and forward them to a ground gateway. Let $a_l(t)$ and $d_l(t)$ denote the numbers of packets received and forwarded by LEO $l$ at time $t$ (throughout paper, a time-slotted system is considered, where $t$ is an integer).

The LEO satellites are equipped with OBP so that the received packets can be stored prior to forwarding. Let $q_l (t)$ represent the length of the queue (i.e., the state of queue) at LEO $l$. Then, we can show that
\be 
q_l (t+1) = (q_l (t) + a_l (t) - d_l (t))^+ ,
    \label{EQ:q_l}
\ee 
where $(x)^+ = \max\{0, x\}$.

\begin{figure}[thb]
\begin{center}
\includegraphics[width=0.6\columnwidth]{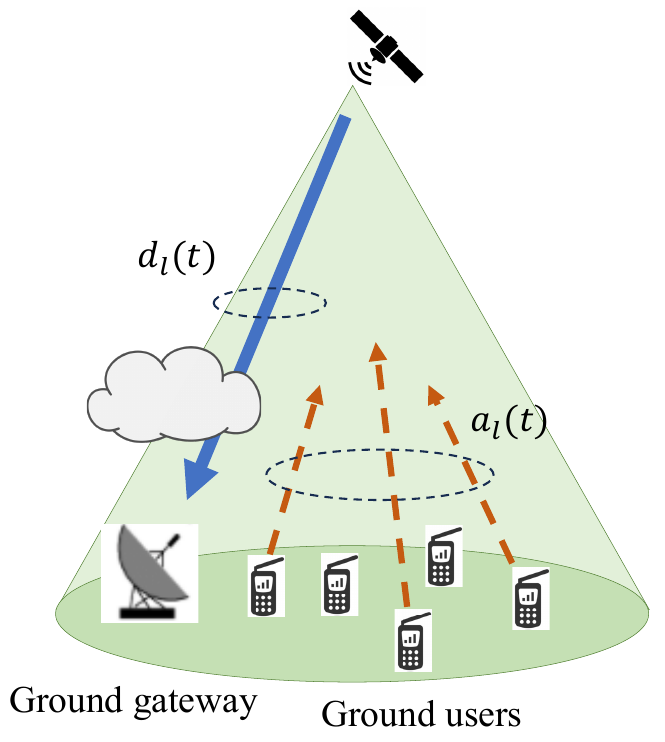} 
\end{center}
\caption{An illustration of satellite communication with ground users and ground gateway.}
        \label{Fig:XLEO}
\end{figure}

While there can be various models for the arrival and departure processes, $a_l(t)$ and $d_l(t)$, respectively, we adopt specific models in this paper for simplicity. Particularly, we assume that $a_l(t)$ follows a Poisson distribution as follows:
\be 
\Pr(a_l (t) = n) = \frac{e^{-\lambda} \lambda^n} {n!} ,
\ n = 0,1,\ldots,
    \label{EQ:a_l}
\ee 
where $\lambda$ represents the average number of active ground users sending data packets within the current footprint or spot-beam. This assumption is valid when the ground users are uniformly deployed over an area according to a homogeneous point Poisson process (PPP), as illustrated in Fig.~\ref{Fig:XLEO}, where ground users (which can be IoT devices \cite{Fraire2019}) are independently active to transmit data packets in each time slot.

For the departure process, we have $d_l(t) = c$, where $c > 0$ is the normalized transmission rate in terms of the number of packets per unit time (here, the unit time is equivalent to the length of a time slot), when reliable transmissions are possible from LEO satellites to ground gateways. Generally, ground gateways are equipped with directional antennas, enabling high transmission reliability from LEO satellites without interference as illustrated in Fig.~\ref{Fig:XLEO}. However, in the event of poor weather conditions, reliable transmissions become difficult, i.e., $d_l (t) = 0$. Thus, the resulting channel for the satellite-ground gateway link can be modeled as an erasure channel.

Note that $a_l(t)$ and $d_l(t)$ are assumed to be uncorrelated, as the footprint of an LEO satellite can be large and weather conditions within the footprint can significantly vary. However, $d_l(t)$ may not be independent and identically distributed (iid), since the transmission is dependent on the local weather of the ground gateway as shown in Fig.~\ref{Fig:XLEO}. 
Thus, we consider a two-state Markov model for $d_l(t)$ as follows:
\begin{align}
p_{ij} = 
\Pr(d_l (t+1) = c_j\,|\, d_l (t) = c_i) , \ i,j \in \{0,1\},
    \label{EQ:d_l}
\end{align}
where $c_0 = 0$ represents the state of poor channel conditions, and $c_1 = c$ represents the state of good channel conditions. Here, let $\alpha = p_{01}$ and $\beta = p_{10}$ denote the transition probabilities from bad-to-good and good-to-bad weather conditions, respectively, as the LEO moves along its orbit.

\section{Performance Analysis of a Single LEO}

We assume that there are no ISL between LEO satellites. Thus, we can consider only one LEO satellite to understand the performance using the state of queue in \eqref{EQ:q_l}.

It is expected to have $\uE[a_l (t)] < \uE[d_l (t)]$ to prevent the queue length from growing indefinitely. Throughout the paper, we assume this condition, i.e., $\uE[a_l (t)] < \uE[d_l (t)]$. From \eqref{EQ:a_l}, we have $\uE[a_l (t)]= \lambda$. For the two-state Markov chain in \eqref{EQ:d_l}, the stationary distribution is given by
$(\pi_0, \pi_1) = \left( \frac{\beta}{\alpha+\beta}, \frac{\alpha}{\alpha+\beta} \right)$,
where $\pi_0 = \Pr(d_l (t) = 0)$ and $\pi_1 = \Pr(d_l (t) = c)$. 
Thus, we assume that 
\be 
\lambda < \uE[c_l (t)] = c \pi_1 = c \frac{\alpha}{\alpha+\beta} .
    \label{EQ:Stability}
\ee 

In addition, for a low buffer overflow probability, we expect to have
\be 
\Pr(q_l (t) > \tau) \le e^{- \theta \tau},
\ee 
where $\tau > 0$ is the threshold of the queue length and $\theta$ is called the quality-of-service (QoS) exponent, which depends on the statistical properties of $a_l (t)$ and $d_l (t)$ \cite{Chang95}.
In particular, it is shown that $\theta$ is the solution of the following equation:
\be 
\Lambda_A (\theta) + \Lambda_D (-\theta) = 0,
    \label{EQ:LL}
\ee 
where $\Lambda_A (\theta)$ and $\Lambda_D (\theta)$ are the logarithmic moment generating functions (LMGF) of $a_l(t)$ and $d_l(t)$, respectively, which are given by
$\Lambda_A (\theta) 
 = \lim_{T \to \infty} \frac{1}{T} \log \uE[ e^{ \theta 
\sum_{t=0}^{T-1} a_l (t) }] =  \log \uE[ e^{ \theta 
a_l (t) }]$ and
$\Lambda_D (\theta) 
 = \lim_{T \to \infty}  \frac{1}{T}  \log \uE[ e^{ \theta 
\sum_{t=0}^{T-1} d_l (t) }]$.
Here, $a_l (t)$ is assumed to be iid. That is, in each time slot, the number of the received packets is independent. From \eqref{EQ:a_l}, we have
$\Lambda_A (\theta) = \lambda (e^\theta - 1)$. 

Recalling that $d_l (t)$ is a two-state Markov chain, from \cite{Ganesh04} \cite{Choi23}, we have
\begin{align} 
& \Lambda_D (\theta) = \cr 
& \frac{1}{\theta} \log 
\left(
\frac{ \bar \alpha + \bar \beta e^{c \theta}
+
\sqrt{(\bar \alpha + \bar \beta e^{c \theta})^2 - 4 
( \bar \beta - \alpha) e^{c \theta}}
}{2}
\right) , \quad \ \ 
\end{align} 
where $\bar \alpha  = p_{00} = 1 - \alpha$ and $\bar \beta  = p_{11} = 1 - \beta$.

Since the size of the buffer is limited, we have $\tau = q_{\rm max}$, where $q_{\rm max}$ represents the buffer size. Consequently, $\Pr(q_l (t) > q_{\rm max})$ represents the packet loss rate due to buffer overflow, which is expected to be sufficiently low for a reliable LEO network. 

\section{Queue-Length Balancing through ISL}

In this section, we consider a LEO satellite network with high-speed ISL, where LEO satellites are interconnected, and discuss queue-length balancing through ISL.

\subsection{High-Speed ISL}

\begin{figure}[thb]
\begin{center}
\includegraphics[width=0.6\columnwidth]{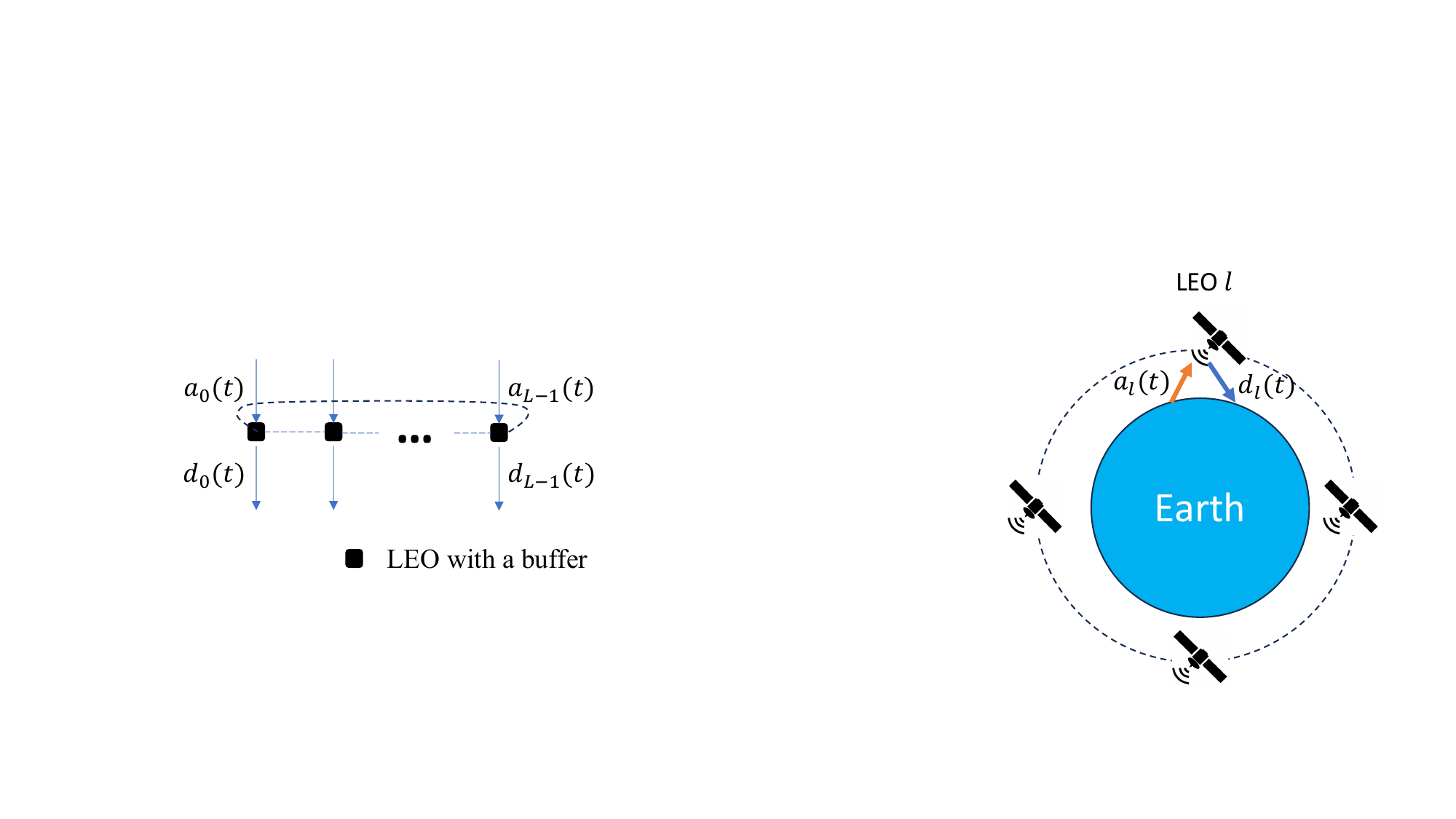} 
\end{center}
\caption{An inter-connected LEO satellite network with individual buffers, where dashed lines represent ISLs.}
        \label{Fig:ISL}
\end{figure}

The transmission rate of laser ISLs can be up to 10 Gbps \cite{Chaudhry21}, which might be much higher than a total transmission rate of ground-satellite links. 
As a result, the LEO satellite network consisting of all LEO satellites in an orbit can be seen as a multiple input multiple output system with distributed queues that are reliably connected via ISLs as illustrated in Fig.~\ref{Fig:ISL}.

Note that the total arrivals and departures, $\sum_l a_l (t)$ and $\sum_l d_l (t)$, respectively, remain unchanged compared to scenarios without ISLs. Thus, throughput may not serve as an appropriate metric for comparison. Instead, the probability of buffer overflow can provide insight into the advantages of employing ISLs.

\subsection{Analysis of QoS  with Equal Queue Lengths}

Thanks to high-speed ISLs, we can consider a virtual queue that can store all the arrival packets. As a result, the state of this virtual queue is updated as follows:
\be 
Q (t+1) = \left(Q (t) +\sum_{l=0}^{l-1} a_l (t) - \sum_{l=0}^{l-1} d_l (t) \right)^+,
    \label{EQ:bQ}
\ee 
where $Q(t)$ represents the sum of the individual LEO's buffer lengths, i.e., $Q(t) = \sum_{l=0}^{l-1} q_l (t)$.
Thus, the normalized queue length of the virtual queue becomes $\bar q (t) = \frac{Q(t)}{L}$. 
Accordingly, the probability of buffer overflow with threshold $\tau$ is given by
\begin{align}
P_{\rm out}  = \Pr( \bar q (t) > \tau) ,
\end{align}
where $\bar q (t+1) = \left(\bar q (t) +
\bar a (t) -  \bar d (t) \right)^+$ from \eqref{EQ:bQ}.
Here, $\bar a(t) = \frac{1}{L} \sum_{l=0}^{L-1} a_l (t)$ and
$\bar d(t) = \frac{1}{L} \sum_{l=0}^{L-1} d_l (t)$.

\begin{mylemma}
Suppose that the arrival and departure processes of each LEO satellite are independent.
Let $\theta^\ast$ be the solution of \eqref{EQ:LL}. Then, the QoS exponent for the virtual queue in \eqref{EQ:bQ}, denoted by $\bar \theta^\ast$, becomes
\be 
\bar \theta^\ast = L \theta^\ast.
    \label{EQ:L1}
\ee 
\end{mylemma}
\begin{IEEEproof} 
Denote by $\Lambda_{\bar A} (\theta)$ and $\Lambda_{\bar D} (\theta)$ the LMGFs of $\bar a_l(t)$ and $\bar d_l(t)$, respectively.
Since
$\Lambda_{\bar A} (\theta) = L \Lambda_A \left(
\frac{\theta}{L} \right)$
and  
$\Lambda_{\bar D} (\theta) = L \Lambda_D \left(
\frac{\theta}{L} \right)$,  \eqref{EQ:LL} becomes
\be
\Lambda_A \left(
\frac{\theta}{L} \right) 
+ \Lambda_D \left(
\frac{\theta}{L} \right) = 0 ,
\ee 
which leads to $\frac{\bar \theta^\ast}{L} = \theta^\ast$, i.e.,
\eqref{EQ:L1}.
\end{IEEEproof}

While the QoS exponent appears to improve with ISLs in \eqref{EQ:L1}, the necessity of a virtual queue to store packets from all $L$ LEO satellites in a central location renders it impractical. Thus, it is necessary to consider how to reallocate packets among the $L$ LEO satellites via ISLs to reduce the probability of buffer overflow at each LEO satellite.

\subsection{Optimal Queue-Length Balancing}

Suppose that each LEO first stores received packets and then distributes them. Thus, each LEO may have the following number of packets to store prior to reallocation:
\be 
s_l (t+1) = q_l (t) + a_l (t) - d_l (t) .
\ee 
Then, optimal reallocation of packets to minimize the maximum number of packets among $L$ LEO satellites can be considered. The resulting problem can be formulated as follows:
\begin{eqnarray}
   & \min_{ \{q_l (t) \} } \max_l \uE[s_l (t+1) \,|\, d_l(t-1)] & \cr
   & \mbox{subject to}\ \sum_{l=0}^{L-1} q_l (t) = Q(t) 
\   \mbox{and}\ q_l (t) \ge 0, \ \forall l .&
    \label{EQ:opt}
\end{eqnarray}
In \eqref{EQ:opt}, the objective function minimizes the maximum expected number of packets stored at any LEO satellite, ensuring a balanced distribution of packet loads across all satellites. By redistributing packets among the LEOs' buffers to equalize their lengths, the optimization minimizes the risk of buffer overflow at any individual LEO.

Note that since $a_l (t)$ and $d_l(t)$ are unknown prior to the reallocation of the stored packets, the mean of $s_l (t+1)$ is used for the objective function in \eqref{EQ:opt}. In addition, because $d_l (t)$ is a Markov chain,  the conditional mean is considered. Then, it can be shown that
\begin{align}
\uE[s_l (t+1) \,|\, d_l(t-1)] 
& = q_l (t) \cr
& \mkern-18mu + \uE[a_l (t)] - \uE[d_l (t)\,|\, d_l (t-1)],\ \ 
    \label{EQ:esd}
\end{align}
and $q_l (t)$ is optimized by reallocating the stored packets among LEO satellites. 

\begin{mylemma} \label{L:2}
Denote by $q_l^\ast (t)$ the optimal solution of the problem in \eqref{EQ:opt}.
Let $\delta = c (\bar \beta - \alpha)$ and $Z =
\sum_{l=0}^{L-1} \indicator(d_l (t-1)=c)$,
where $\indicator(X)$ represents the indicator function, which is 1 when $X$ is true and 0 otherwise.    
Then, there are 3 different cases for the optimal solution as follows:
\begin{itemize}
\item Case I: If $\delta > \frac{Q(t)}{Z}$,
\be
q^\ast_l (t)
= \left\{
\begin{array}{ll}
0, & \mbox{if $d_l (t-1) = 0$} \cr
\frac{Q(t)}{Z}, & \mbox{if $d_l (t-1) = c$} \cr
\end{array}
\right.
    \label{EQ:C1}
\ee 
\item Case II: If $\delta < -\frac{Q(t)}{L-Z}$,
\be
q^\ast_l (t)
= \left\{
\begin{array}{ll}
\frac{Q(t)}{L-Z}, & \mbox{if $d_l (t-1) = 0$} \cr
0, & \mbox{if $d_l (t-1) = c$} \cr
\end{array}
\right.
    \label{EQ:C2}
\ee 
\item Case III: If $-\frac{Q(t)}{L-Z}\le \delta \le \frac{Q(t)}{Z}$,
\be
q^\ast_l (t)
= \left\{
\begin{array}{ll}
\frac{Q(t) - Z \delta}{L}, & \mbox{if $d_l (t-1) = 0$} \cr
\frac{Q(t) + (L-Z) \delta}{L}, & \mbox{if $d_l (t-1) = c$.} \cr
\end{array}
\right.
    \label{EQ:C3}
\ee    
\end{itemize}
\end{mylemma}
\begin{IEEEproof}
For convenience, let $Q = Q(t)$ and $q_l = q_l (t)$. For each LEO, the right-hand side (RHS) term in \eqref{EQ:esd} has one of the following two possible values:
\be 
{\rm RHS} = \left\{
\begin{array}{ll}
q_l + \lambda - c \alpha, & \mbox{if $d_l (t-1) = 0$} \cr
q_l + \lambda - c \bar \beta, & \mbox{if $d_l (t-1) = c$.} \cr
\end{array}
\right.
    \label{EQ:RHS}
\ee 
As a result, we can claim that $q_l^\ast \in \{x, y\}$, where $x$ and $y$ represent the solutions when $d_l(t-1) = 0$ and $d_l (t-1) = c$, respectively, while $x, y \ge 0$.
Due to the constraint that $\sum_{l=0}^{L-1} q_l = Q$, we have
\be 
x (L-Z) + y Z = Q.
    \label{EQ:first}
\ee 

From \eqref{EQ:RHS}, suppose that
$x + \lambda - c\alpha \ge y + \lambda - c \bar \beta$
or 
\be 
x \ge y - \delta.
    \label{EQ:xyd}
\ee 
Then, we have
\be 
\max_l \uE[s_l (t+1) \,|\, d_l(t-1)] = x + \lambda - c\alpha ,
\ee 
which is to be minimized. Thus, the equality in \eqref{EQ:xyd} is to be achieved, i.e.,
\be 
x = y - \delta.
    \label{EQ:second}
\ee 
We can find $(x,y)$ by solving \eqref{EQ:first} and \eqref{EQ:second}. However, since $x$ and $y$ are non-negative, there are 3 different cases. For example, if $\delta > \frac{Q}{Z}$, the value of $x$ satisfying \eqref{EQ:first} and \eqref{EQ:second} becomes negative. Thus, in this case, with $x = 0$, we have $y = \frac{Q}{Z}$ to meet $Q =\sum_l q_l = y Z$, which is given in \eqref{EQ:C1}. In addition, if $\delta < -\frac{Q}{L-Z}$, $y$ becomes negative. Again, in this case, we need to choose $y = 0$, and then $x = \frac{Q}{L-Z}$ to meet $Q =\sum_l q_l = x (L- Z)$, which is given in \eqref{EQ:C2}. Finally, when $-\frac{Q}{L-Z}\le \delta \le \frac{Q}{Z}$, $(x,y)$ can be found by solving \eqref{EQ:first} and \eqref{EQ:second}, which is given in \eqref{EQ:C3}.

Note that we can have the same result when $x + \lambda - c\alpha \le y + \lambda - c \bar \beta$ is assumed. This completes the proof.
\end{IEEEproof}

To understand the allocation according to Lemma~\ref{L:2}, let us assume that $p_{11} = \bar \beta \gg p_{01} = \alpha$. Then, $\frac{Q(t)}{Z}$ can be interpreted as the predicted number of reallocated packets per LEO likely to be in a good state. From this, the condition of Case I, $\delta > \frac{Q(t)}{Z}$, which can be rewritten as: 
$c p_{11} > c p_{01} + \frac{Q(t)}{Z}$,
implies that the expected number of packets to be transmitted by a LEO in a good state is greater than the sum of the predicted number of packets per LEO and the number of packets transmitted by a LEO transitioning from a bad to a good state. In such a case, the LEO satellites with bad states in the previous time slot will have zero packets, while each of those with good states in the previous time slot will have $\frac{Q(t)}{Z}$ packets as shown in \eqref{EQ:C1}.

We now assume that $\alpha = p_{01} \gg \bar \beta = p_{11}$. In this case, $\frac{Q(t)}{L-Z}$ (rather than $\frac{Q(t)}{Z}$) becomes the predicted number of reallocated packets per LEO likely to be in a good state. Then, the condition of Case II is rewritten as:
$c p_{01} > c p_{11} + \frac{Q(t)}{L-Z}$,
which means that the expected number of packets to be transmitted by a LEO in a good state from a bad state is greater than the sum of the predicted number of packets per LEO and the number of packets transmitted by a LEO staying in a good state. The corresponding allocation of packets is given in \eqref{EQ:C2}.

Alternatively, the packets can be optimally redistributed according to \eqref{EQ:C3}. The resulting approach that allocates the packets according to Lemma~\ref{L:2} is referred to as the minimum queue-length allocation via ISL (MQLA-ISL).

\section{Simulation Results}

In this section, we present simulation results of the probability of buffer overflow when the arrival processes follow a Poisson distribution as in \eqref{EQ:a_l}, and the channel condition of satellite-ground gateway links can be modeled as a two-state Markov chain as described in \eqref{EQ:d_l}.

\begin{figure}[thb]
\begin{center}
\includegraphics[width=0.67\columnwidth]{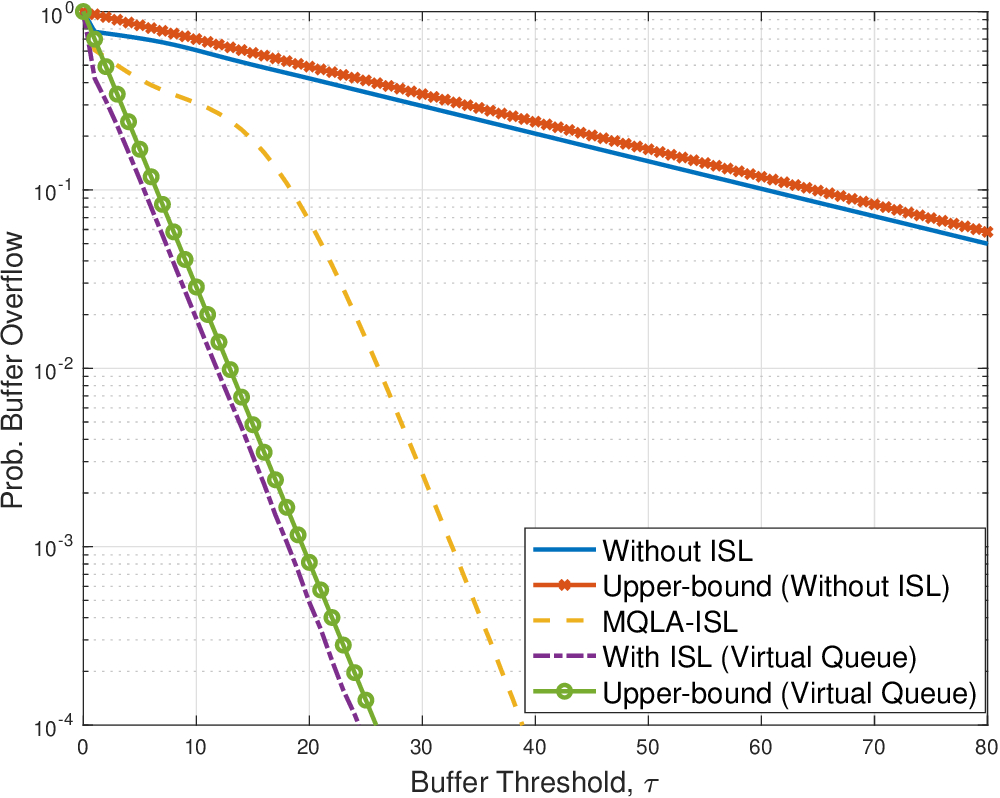} 
\end{center}
\caption{The probability of buffer overflow versus threshold, $\tau$, when $(\alpha, \beta) = (0.7, 0.3)$, $\lambda = 10$,
$c = 16$, and $L = 10$.}
        \label{Fig:OQ}
\end{figure}

Fig.~\ref{Fig:OQ} shows the probability of buffer overflow as a function of the threshold, $\tau$, when $(\alpha, \beta) = (0.7, 0.3)$, $\lambda = 10$, $c = 16$, and $L = 10$. It demonstrates that the probability of buffer overflow in the proposed method, i.e., MQLA-ISL, is lower than that of the system without ISL. In addition, as $\tau$ increases, the probability approaches that of the ideal system with a virtual queue. 
An important observation is that MQLA-ISL requires a buffer size of $q_{\rm max} = \tau = 40$ to achieve a probability of buffer overflow of $10^{-4}$. However, without ISLs, this probability of buffer overflow can be achieved with a very large buffer size (which is approximately $259$\footnote{The QoS exponent satisfying \eqref{EQ:LL} for the given values of the parameters is $\theta =0.0356$ when no ISLs are used. Thus, a probability of buffer overflow of $10^{-4}$ can be achieved when $q_{\rm max} = 259$.}, although it is not shown in Fig.~\ref{Fig:OQ}). This means that the size of buffer can be significantly reduced when ISLs are employed in conjunction with the optimal packet reallocation (i.e., MQLA-ISL).

\begin{figure}[!t]\vspace{-1em}
\begin{center} \subfigure[$q_{\rm max} = 15$]{\label{fig 1 ax}\includegraphics[width=0.49\columnwidth]{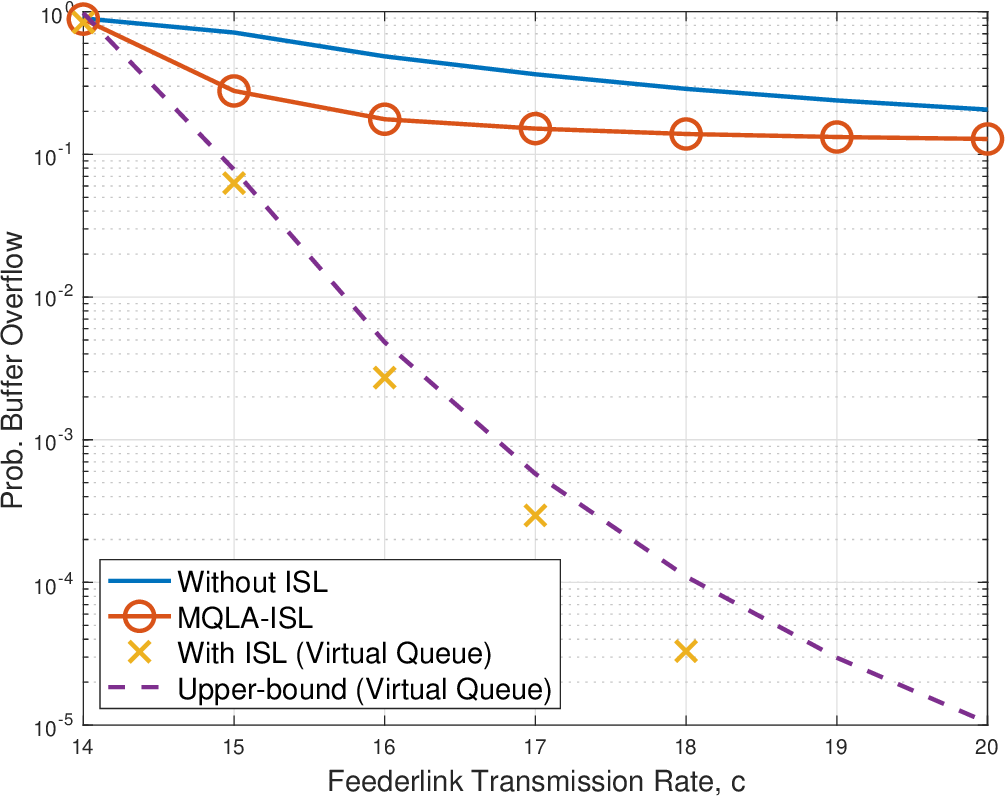}}
\subfigure[$q_{\rm max} = 30$]{\label{fig 1 bx}\includegraphics[width=0.49\columnwidth]{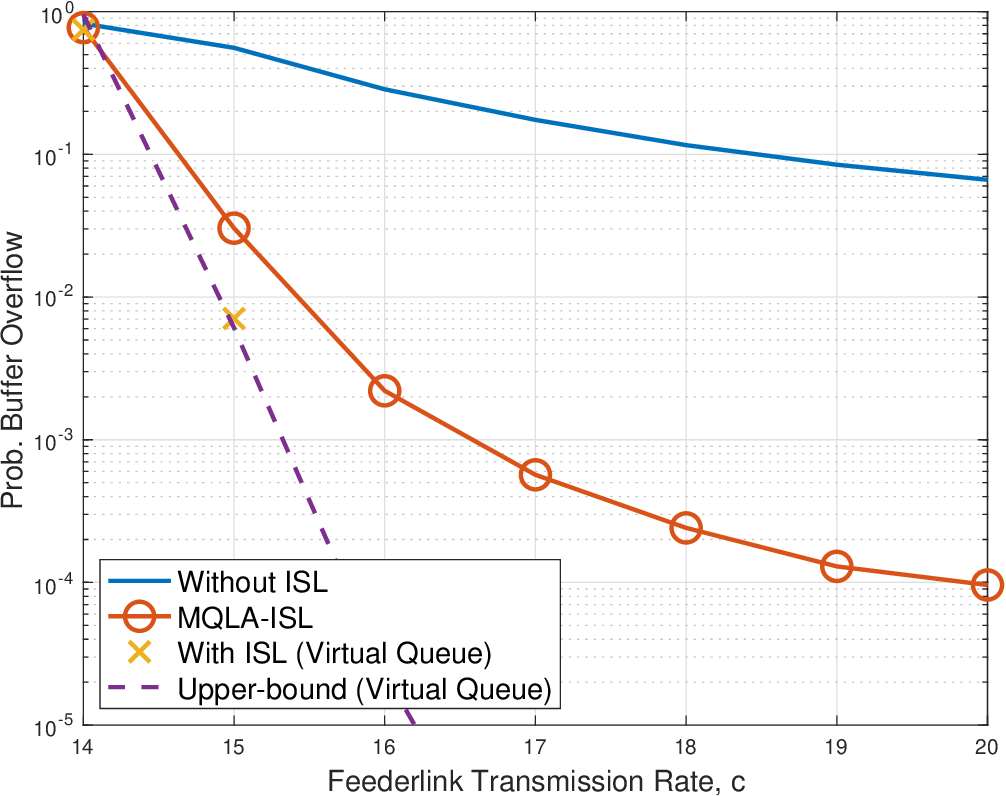}} \vspace{-1em}
\end{center}
\caption{Probability of buffer overflow versus $c$ when
$(\alpha, \beta) = (0.7, 0.3)$, $\lambda = 10$, and
$L = 10$: (a) $q_{\rm max} = 15$; (b) $q_{\rm max} = 30$.}
    \label{Fig:plt_cs}
\end{figure}

In Fig.~\ref{Fig:plt_cs}, the probability of buffer overflow is shown as a function of the transmission rate of the feeder link, $c$, when $(\alpha, \beta) = (0.7, 0.3)$, $\lambda = 10$, $L = 10$, and $q_{\rm max} \in \{15, 30\}$. Note that $c$ has to be greater than $\lambda \frac{\alpha+\beta}{\alpha} = 14.2857$ from \eqref{EQ:Stability}. Thus, when $c = 14$, the probability of buffer overflow becomes 1 regardless of ISL, while it decreases with $c$. It is noteworthy that with a larger $q_{\rm max}$, the probability of buffer overflow in MQLA-ISL is closer to that of the ideal system with a virtual queue, which can be seen by comparing Figs.~\ref{Fig:plt_cs} (a) and (b).

\begin{figure}[!t]\vspace{-1em}
\begin{center} \subfigure[$q_{\rm max} = 15$]{\label{fig 1 ax}\includegraphics[width=0.49\columnwidth]{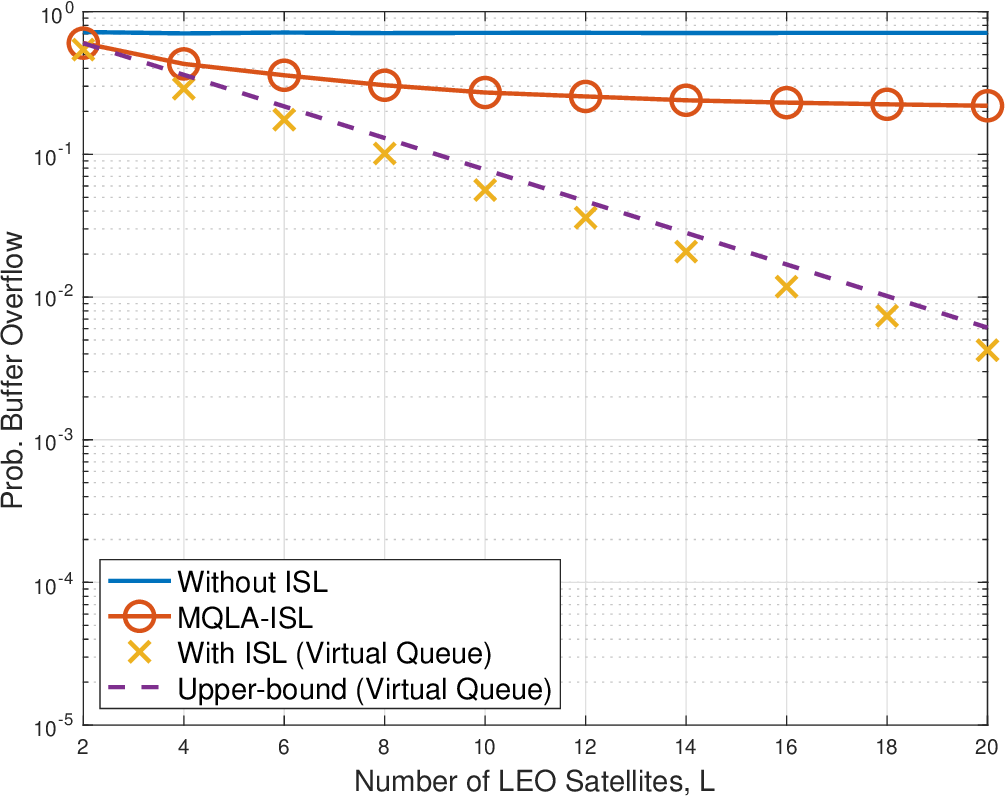}}
\subfigure[$q_{\rm max} = 30$]{\label{fig 1 bx}\includegraphics[width=0.49\columnwidth]{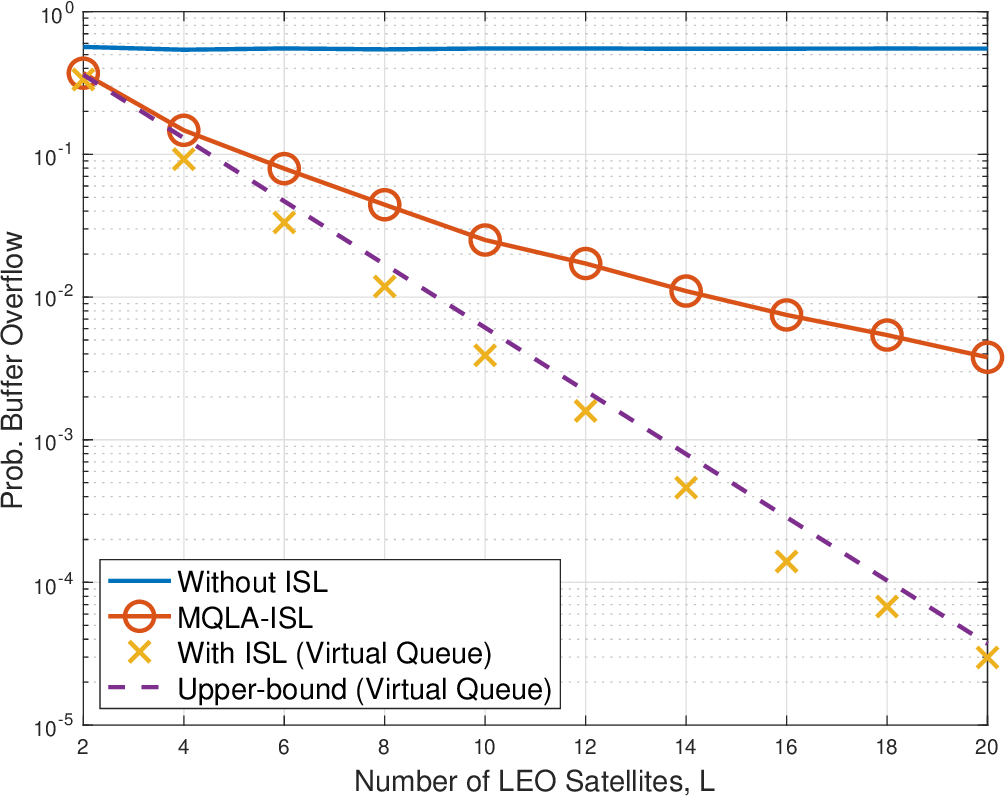}} \vspace{-1em}
\end{center}
\caption{Probability of buffer overflow versus $L$ when
$(\alpha, \beta) = (0.7, 0.3)$, $\lambda = 10$, and
$c = 16$: (a) $q_{\rm max} = 15$; (b) $q_{\rm max} = 30$.}
    \label{Fig:plt_L}
\end{figure}

In Fig.~\ref{Fig:plt_L}, the impact of the number of LEO satellites, $L$, on the probability of buffer overflow is shown when $(\alpha, \beta) = (0.7, 0.3)$, $\lambda = 10$, $c = 16$, and $q_{\rm max} \in \{15, 30\}$. It is demonstrated that the probability of buffer overflow in MQLA-ISL decreases with increasing $L$, akin to the behavior observed in the ideal system with a virtual queue, especially when $q_{\rm max}$ is large, while the system without ISL does not exhibit such behavior. This illustrates that MQLA-ISL enables the reduction of buffer size (or storage) while maintaining a certain level of overflow probability.

\begin{figure}[!t]\vspace{-1em}
\begin{center} \subfigure[overflow prob. vs. $\alpha$]{\label{fig 1 ax}\includegraphics[width=0.49\columnwidth]{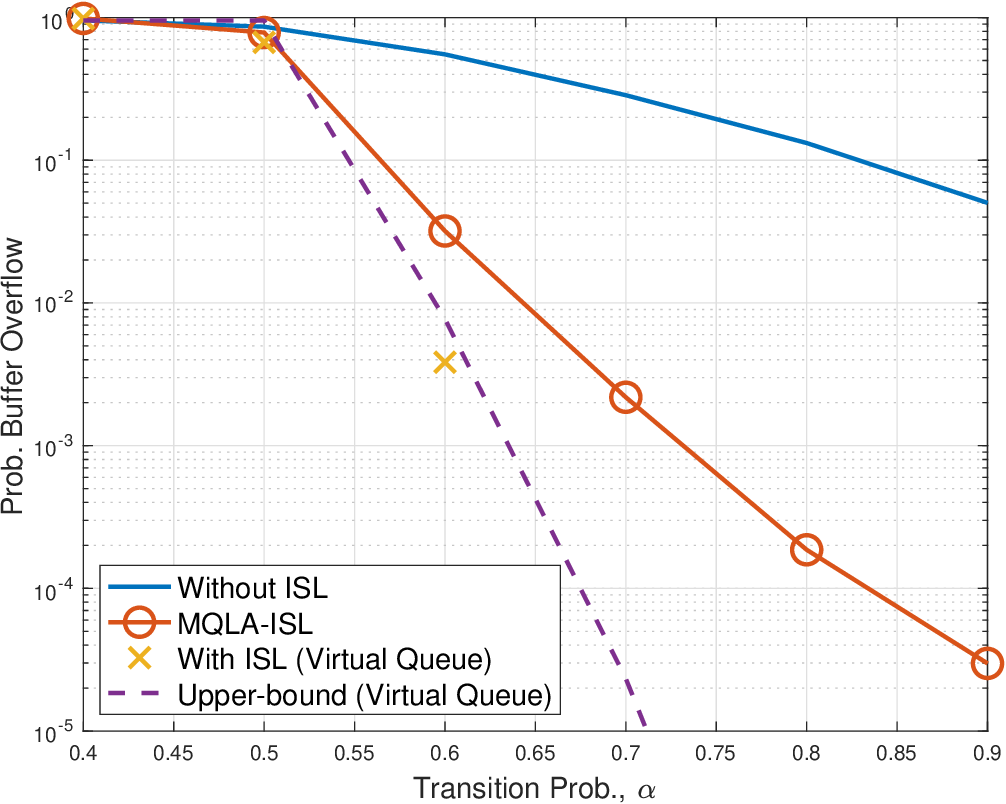}}
\subfigure[overflow prob. vs. $\beta$]{\label{fig 1 bx}\includegraphics[width=0.49\columnwidth]{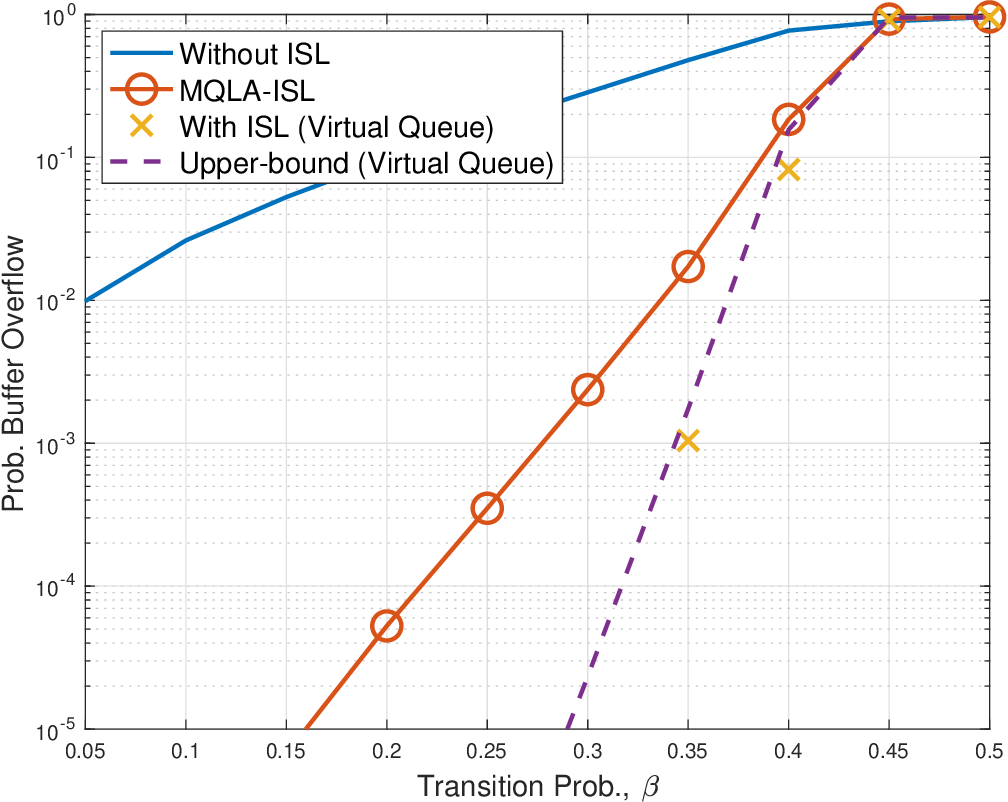}} \vspace{-1em}
\end{center}
\caption{Probability of buffer overflow versus $\alpha$ and $\beta$ with
$q_{\rm max} = 30$, $\lambda = 10$, $c=16$, and
$L = 10$: (a) varying $\alpha$ ($\beta =0.3$); (b) varying  $\beta$ ($\alpha =0.7$).}
\label{Fig:plt_as}
\end{figure}

In Fig.~\ref{Fig:plt_as}, the impact of transition probabilities ($\alpha$ and $\beta$) on the probability of buffer overflow is demonstrated when  $\lambda = 10$, $c = 16$, $L = 10$, and $q_{\rm max} = 30$. As $\alpha$ increases, $\pi_1$ increases, indicating a higher average transmission rate from LEO satellites to ground gateways.
Consequently, in Fig.~\ref{Fig:plt_as} (a), the probability of buffer overflow decreases with increasing $\alpha$. In addition, as shown in Fig.~\ref{Fig:plt_as} (b), the probability of buffer overflow decreases with decreasing $\beta$ as the channel state remains in a good state more often.
Again, it is observed that the behavior of the probability of buffer overflow in MQLA-ISL closely resembles that of the ideal system with a virtual queue, particularly for large $q_{\rm max}$.

\section{Conclusions}

In this paper, we examined the impact of ISLs on the probability of buffer overflow in LEO satellite networks with store-and-forward transmission. Our analysis underscored the pivotal role of ISLs in enhancing network reliability. By examining finite buffer sizes, we emphasized the critical importance of maintaining minimal buffer overflow probabilities to ensure reliable network performance and potentially reduce buffer sizes while meeting a specified target probability of overflow using ISL. Furthermore, we derived a practical optimal packet reallocation strategy through ISL, which can minimize the probability of buffer overflow, thereby contributing to a more reliable LEO network.

\bibliographystyle{ieeetr}
\bibliography{LEO}

\end{document}